\documentclass[twocolumn]{svjour3}

\smartqed  

\usepackage{graphicx}
\usepackage{layout}
\usepackage{textcomp}
\usepackage[toc,page,titletoc]{appendix}

\usepackage{graphics}
\usepackage{footnote}
\usepackage{tabularx,ragged2e,booktabs}
\usepackage{array}
\usepackage{natbib}
\usepackage{amsmath}
\usepackage{amssymb}
\usepackage{color}
\usepackage{url}

\begin{document}

\title{A Database of Phase Calibration Sources and their Radio Spectra for the Giant Metrewave Radio Telescope}

\titlerunning{GMRT Calibrator Manual}        

\author{Dharam V. Lal \and Shilpa S. Dubal \and Sachin S. Sherkar}

\authorrunning{Lal, D. V.} 

\institute{Dharam V. Lal, Shilpa S. Dubal \& Sachin S. Sherkar \at
              National Centre for Radio Astrophysics (NCRA--TIFR), Post Box 3, Ganeshkhind P.O. Pune 411007. \\
              \email{dharam@ncra.tifr.res.in}           
}


\maketitle

\begin{abstract}

We are pursuing a project to build a database of phase calibration sources suitable
for Giant Metrewave Radio Telescope (GMRT).
Here we present the first release of 45 low frequency calibration sources
at 235 MHz and 610 MHz.
These calibration sources are broadly divided into quasars, radio galaxies
and unidentified sources.
We provide their flux densities, models for calibration sources, ($u,v$) plots,
final deconvolved restored maps and \textsc{clean}-component lists/files
for use in the Astronomical Image Processing System (\textsc{aips}) and
the Common Astronomy Software Applications (\textsc{casa}).
We also assign a quality factor to each of the calibration sources.
These data products are made available online through the GMRT observatory
website.
In addition we find that (i) these 45 low frequency calibration sources are
uniformly distributed in the sky and future efforts to increase the size of
the database should populate the sky further,
(ii) spectra of these calibration sources are about equally divided
between straight, curved and complex shapes,
(iii) quasars tend to exhibit flatter radio spectra as compared to the
radio galaxies or the unidentified sources, 
(iv) quasars are also known to be radio variable and hence possibly show complex spectra
more frequently, and
(v) radio galaxies tend to have steeper spectra, which
are possibly due to the large redshifts of distant galaxies
causing the shift of spectrum to lower frequencies.

\keywords{surveys -- techniques \and interferometric -- techniques \and high angular resolution -- radio continuum \and general}
\end{abstract}

\section{Introduction}

The Giant Metrewave Radio telescope
\citep[GMRT,][]{Swarupetal} is the most sensitive radio telescope in the world
that is capable of operating at low
radio frequencies from 150 MHz to 1450 MHz. In order to obtain reliable information about target
astronomical sources, radio sources whose structure and flux density are known
{\it a~priori} are routinely observed to determine antenna based calibration
solutions. Any errors in the models used for these calibration sources reflect
in errors in the antenna based calibration solutions, especially in the phase,
and limit the quality of the radio images obtained.
GMRT users have traditionally been using phase calibration sources from
the Very Large Array (VLA) \citet{vla-cal-manual},
which is biased towards observations optimized for higher frequencies.
In this paper we present a new database which is intended to facilitate science
operations at the GMRT.
The database presenting all products, i.e.,
their flux densities, models, ($u,v$) plots, final deconvolved restored maps and
\textsc{clean}-component lists/files \citep{Cornwelletal,hogbom1974},
is available at the `Observing Help' web-page of the GMRT observatory.

In this paper, our motivations for such a database are discussed
in Sect.~\ref{motiv}.
The sample is presented in Sect.~\ref{sample} and its archival data and
data reduction are detailed in Sect.~\ref{obs-data} and Sect.~\ref{data-red}, respectively.
The results from the database, i.e., 
their quality standards, sky distribution,
comparison of images made using image model from the database and made in a routine manner and radio properties are discussed
in Sect.~\ref{qual-fac}, Sect.~\ref{sky_dist}, Sect.~\ref{comparison} and
Sect.~\ref{int-rad-spec}, respectively.
Finally, we provide a summary in Sect.~\ref{summary}.

\section{Motivations}
\label{motiv}

The process of removing effects of phase and amplitude corruption from
interferometer data is called calibration \citep{FP99}.
Several complexities and levels in the calibration of
interferometric instruments exist,
for example, errors due to
(i) changes in the length of transmission system or receiver sensitivity and the electronics,
(ii) inaccurate antenna positions (i.e., geometric errors), which are typically
known to $\approx$ cm--m,
and (iii) changes in the visibility induced by the atmosphere, including ionosphere
\citep{BC1995}.
The observed visibility phase is defined as the relative phase of
the electric fields measured simultaneously at two points along the
wavefront from the radio sources.  The atmospheric effects on this
phase can be important on timescales ranging 
from less than a minute to hours or more.

Mathematically, the error in observed visibility phase $\phi_{\rm obs}(t,~\nu)$
from the correlator can be expressed as
\begin{eqnarray}\nonumber
\begin{split}
\phi_{\rm obs}(t,~\nu) = \phi_{\rm true}(t,~\nu) + \phi_{\rm inst}(t,~\nu) \\
 + \phi_{\rm geom}(t,~\nu) + \phi_{\rm atmos}(t,~\nu)
\end{split}
\end{eqnarray}
where $\phi_{\rm true}(t,~\nu)$ is the true visibility phase,
$\phi_{\rm inst}(t,~\nu)$ is the sum of all instrumental phase errors
at the two antennas and is possibly due to propagation through the antenna optics
and electronics before the sampled electric field is digitized,
$\phi_{\rm geom}(t,~\nu)$ is the phase error due to geometrical
errors in the delay model and $\phi_{\rm atmos}(t,~\nu)$ represent the effects of
different atmospheric delays above each antenna.
Note that both the ionospheric and the tropospheric \citep{Thompsonetal,CH1999} errors
are included in phase errors due to the atmosphere.
The dominant sources of errors
at low frequencies ($\lesssim$~1 GHz) are refraction,
propagation delay and Faraday rotation caused by the ionosphere,
while at high frequencies ($\gtrsim$~1 GHz) they are scattering,
effects of water vapor and propagation delay caused by the troposphere \citep{Thompsonetal}.
Instrumental, $\phi_{\rm inst}(t,~\nu)$ and geometrical, $\phi_{\rm geom}(t,~\nu)$
errors can be reduced to acceptable levels by good design,
but corrections for changes in the visibility due to the atmosphere are not possible,
since these changes affect visibility phases.
It is well known that the ionosphere severely limits the phase
stability of radio interferometers at low radio frequencies \citep{Intemaetal}.
In practice, if $\phi_{\rm obs}(t,~\nu)$
varies by a full turn over time or frequency-channel,
it limits an observer's ability to integrate data over time and frequency.
Several techniques are available that use interferometry to overcome the
limits imposed by the atmosphere.
The technique of making observations of a bright nearby point source,
called the `phase' calibration source, frequently interspersed with the target source
is called phase referencing \citep{BC1995}.
This technique of phase referencing has been used successfully to allow
correction of the fast temporal changes of relative phases of
individual antennas of the interferometer.

Methodologically, consider the nodding style,
where observations at time $t_2$ of the target source are interleaved
between calibration observations at $t_1$ and $t_3$ of the phase calibration source.
The observed phases for calibration source and target source, $\phi_{\rm obs}^{\rm c}$ and $\phi_{\rm obs}^{\rm t}$, respectively, will be
\begin{eqnarray}\nonumber
\begin{split}
\phi_{\rm obs}^{\rm c}(t_1,~\nu) = \phi_{\rm true}^{\rm c}(t_1,~\nu),~\nu + \phi_{\rm inst}^{c}(t_1,~\nu) \\
+ \phi_{\rm geom}^{c}(t_1,~\nu) + \phi_{\rm atmos}^{c}(t_1,~\nu),
\end{split}
\end{eqnarray}

\begin{eqnarray}\nonumber
\begin{split}
\phi_{\rm obs}^{\rm t}(t_2,~\nu) = \phi_{\rm true}^{\rm t}(t_2,~\nu) + \phi_{\rm inst}^{t}(t_2,~\nu) \\
+ \phi_{\rm geom}^{t}(t_2,~\nu) + \phi_{\rm atmos}^{t}(t_2,~\nu),
\end{split}
\end{eqnarray}

\begin{eqnarray}\nonumber
\begin{split}
\phi_{\rm obs}^{\rm c}(t_3,~\nu) = \phi_{\rm true}^{\rm c}(t_3,~\nu) + \phi_{\rm inst}^{c}(t_3,~\nu) \\
+ \phi_{\rm geom}^{c}(t_3,~\nu) +
                    \phi_{\rm atmos}^{c}(t_3,~\nu).
\end{split}
\end{eqnarray}
Next let us represent the interpolated phase of calibration source at time $t_2$, by
\begin{eqnarray}\nonumber
\begin{split}
\phi_{\rm obs}^{\prime~{\rm c}}(t_2,~\nu) = \phi_{\rm true}^{\prime~{\rm c}}(t_2,~\nu) + \phi_{\rm inst}^{\prime~{\rm c}}(t_2,~\nu) \\
+ \phi_{\rm geom}^{\prime~{\rm c}}(t_2,~\nu) +
                    \phi_{\rm atmos}^{\prime~{\rm c}}(t_2,~\nu)
\end{split}
\end{eqnarray}
where the interpolated quantities are indicated by primes.
The difference between the phases of the target source and
the interpolated calibration source will be
\begin{eqnarray}\nonumber
\begin{split}
\phi_{\rm obs}^{\rm t} - \phi_{\rm obs}^{\prime~{\rm c}} = (\phi_{\rm true}^{\rm t} - \phi_{\rm true}^{\prime~{\rm c}}) +
 (\phi_{\rm inst}^{\rm t} - \phi_{\rm inst}^{\prime~{\rm c}}) \\
+ (\phi_{\rm geom}^{t} - \phi_{\rm geom}^{\prime~{\rm c}}) +
                   (\phi_{\rm atmos}^{t} - \phi_{\rm atmos}^{\prime~{\rm c}}).
\end{split}
\end{eqnarray}
Here, we have ignored the $t_2$ and $\nu$ notations.
Also, we have determined the telescope phase corrections using the calibration source
and applied to the target source using a suitable interpolation function.
Now we make three reasonable assumptions:
(i) The target source and the calibration source are located roughly in the
same region of the sky, or in other words corrections for one source also apply to the other,
so that
$$
\phi_{\rm atmos}^{t} - \phi_{\rm atmos}^{\prime~{\rm c}} = 0.
$$
This is called the `isoplanarity' assumption.
(ii) If there exist instrumental errors,
they would be the same for the target source and the phase calibration source
due to the design of the instrument,
$$
\phi_{\rm inst}^{\rm t} - \phi_{\rm inst}^{\prime~{\rm c}} = 0,
$$
and
(iii) if there exist
geometrical errors, e.g., antenna position errors,
they, too will be the same for two sources
with a small separation between target source and calibration source,
$$
\phi_{\rm geom}^{t} - \phi_{\rm geom}^{\prime~{\rm c}} = 0.
$$
Thus we obtain,
$$
\phi_{\rm obs}^{\rm t} - \phi_{\rm obs}^{\prime~{\rm c}} = (\phi_{\rm true}^{\rm t} - \phi_{\rm true}^{\prime~{\rm c}}).
$$
Furthermore, either the calibration source is bright, point-like and the only source in 
the field-of-view or we \textit{a~priori} know the model of the field-of-view containing
the calibration source, then we can assume
$$
\phi_{\rm true}^{\prime~{\rm c}} = 0.
$$
The phase-referenced difference phase then contains only information about the
target source structure, and the positions of the target source and calibration source.
Fourier inverting and deconvolving the `phase-calibrated' target source data
should provide us with a good image of the target source.

\section{Sample} \label{sample}

The VLA varies its angular resolution through movement of its antennas,
and hence has four basic antenna configurations, A, B, C and D,
whose scales vary by the ratios 35.5:10.8:3.28:1 \citep{vla-status}.
GMRT, unlike the VLA, has a hybrid configuration, with 14 antennas in a central
array and the remaining 16 antennas spread across three arms of a `Y' \citep{Swarupetal}.
The central array yields baselines of $\lesssim$~1~km
and the longer baselines of the `Y' yield baselines up to a maximum length of $\sim$~25~km,  
i.e., ($u,v$) coverages out to $\sim$ 0.8 k$\lambda$ and $\sim$ 20 k$\lambda$,
respectively at 235 MHz.
In other words, a single observation with the GMRT yields information
from small to large  angular scales, which
in the case of the VLA must be provided by using
several of its antenna configurations.
The quality of VLA calibration sources
varies with frequency and configuration \citep{vla-status}.
Their absolute position errors range from $<$0.002~arcsec
\citep[code `A',][]{vla-status} to $>$0.15~arcsec \citep[code `T',][]{vla-status}.
Correspondingly the phase calibration errors have several levels,
from as small as $<$3\% errors (P-class) or 10\% errors (S-class)
to confused or to unknown or inappropriate.
We selected a list of target calibration sources from the
VLA \citet{vla-cal-manual} which met the following criteria:
The source should be bright, i.e., have flux density at 20~cm, S$_{\rm 20~cm}$ $>$ 0.5 Jy and
(i) should be P-class at A-array and B-array VLA configurations, and
(ii) either P-class or S-class at C-array and D-array VLA configurations.
These criteria provided us with 121 sources, whose positional
uncertainties in right ascension and declination are $<$~0.15~arcsec \citep{vla-status}.
We used this list of 121 phase calibration sources
to look for archival GMRT data and found observations for 45 (34\%) sources
listed in Table~\ref{obs-log} at 235 MHz and 610 MHz.
New observations were made only for a handful of sources when the archival data were badly affected
by radio frequency interference (RFI) or challenging ionospheric conditions;
otherwise no new GMRT observations were made for this project.

These phase calibration sources are broadly divided among three categories,
(i) quasars, (ii) giga-hertz peaked sources, compact steep-spectrum sources
and radio galaxies, which we refer indiscriminately as radio galaxies in
what follows, and
(iii) unidentified sources (see also Table~\ref{obs-log}).
Below we present data and its analysis (Sect.~\ref{obs-data} and~\ref{data-red})
of first release of these 45 sources, forming our database of phase calibration
sources.

\section{Archival Data}
\label{obs-data}

\begin{table*}
\centering
\caption{The observing log for phase calibration sources.
The columns are as follows:
(1) calibration source name. which encodes the J2000 position;
(2) observing date; and
(3) flux density and bandpass calibrator.
Columns (4), (5) and (6), and (7), (8) and (9), respectively are for 235 MHz and 610 MHz data,
where these columns are as follows:
(4, 7) nominal and effective bandwidth;
(5, 8) integration time; and
(6, 9) FWHM of the elliptical Gaussian restoring beam and position angle (P.A.).}
\begin{tabular}{l|cl|clcr|clcr}
\hline
 & & & & & & & & & & \\ [-0.3cm]
 & & & \multicolumn{4}{c|}{235 MHz} & \multicolumn{4}{c}{610 MHz} \\
       \cline{4-11} \\ [-0.3cm]
Calibrator & \multicolumn{2}{l|}{Obs-date Flux density} &\multicolumn{2}{c}{$\Delta\nu$~~~~~t$_{\rm int}$}&\multicolumn{2}{l|}{Synthesized, P.A.}  & \multicolumn{2}{c}{$\Delta\nu$~~~~~t$_{\rm int}$} &\multicolumn{2}{l}{Synthesized, P.A.} \\       
(J2000)    & \multicolumn{2}{r|}{calibrator}      &      &  &   beam       &         &  &     & beam                &  \multicolumn{1}{c}{} \\                               
 & & & \multicolumn{2}{c}{(MHz)~~~(min)} & \multicolumn{2}{c|}{(arcsec$^2$, deg)} & \multicolumn{2}{c}{(MHz)~~~(min)} & \multicolumn{2}{c}{(arcsec$^2$, deg)} \\
 (1) & (2) & (3) & \multicolumn{2}{c}{(4)~~~~~~~(5)} & \multicolumn{2}{c|}{(6)} & \multicolumn{2}{c}{(7)~~~~~~~(8)} & \multicolumn{2}{c}{(9)} \\
\hline
 \multicolumn{11}{l}{Radio galaxies} \\
0010$-$418 & 2013-07-05 & 3C48  & 16 / 15.2 & 67 & \multicolumn{2}{c|}{21.5 $\times$ 10.3,    21.6} & 32 / 31.2 & 85  & \multicolumn{2}{c}{14.3 $\times$ 6.5,    14.8} \\
0029$+$349 & 2006-07-10 & 3C48  & 16 / 14.9 & 33 & \multicolumn{2}{c|}{13.2 $\times$  9.9,    41.9} & 32 / 30.8 & 33  & \multicolumn{2}{c}{ 5.4 $\times$ 4.1,    50.3} \\
0119$+$321 & 2007-08-17 & 3C48  &  6 /  5.4 & 58 & \multicolumn{2}{c|}{14.1 $\times$  9.7,    71.7} & 32 / 30.8 & 58  & \multicolumn{2}{c}{ 6.0 $\times$ 4.7,    82.5} \\
0410$+$769 & 2005-01-22 & 3C147 &  6 /  4.9 &167 & \multicolumn{2}{c|}{18.0 $\times$ 14.6, $-$86.3} & 16 / 15.0 &173  & \multicolumn{2}{c}{10.8 $\times$ 5.8, $-$68.8} \\
0745$+$101 & 2011-11-17 & 3C48  &           &    & \multicolumn{2}{c|}{                           } & 16 / 15.2 & 89  & \multicolumn{2}{c}{ 7.1 $\times$ 6.1,    37.2} \\
0943$-$083 & 2008-12-05 & 3C48  &           &    & \multicolumn{2}{c|}{                           } & 16 / 15.2 & 71  & \multicolumn{2}{c}{ 5.3 $\times$ 4.8,    27.8} \\
1130$-$148 & 2009-06-06 & 3C286 &  6 /  5.2 & 53 & \multicolumn{2}{c|}{16.3 $\times$ 10.8,    21.3} & 32 / 31.4 & 53  & \multicolumn{2}{c}{ 7.1 $\times$ 4.6,    21.2} \\
1313$+$675 & 2011-02-12 & 3C147 &  6 /  4.9 & 48 & \multicolumn{2}{c|}{16.9 $\times$ 13.1, $-$53.2} & 32 / 31.0 & 48  & \multicolumn{2}{c}{ 6.5 $\times$ 5.4, $-$51.8} \\
1400$+$621 & 2004-01-02 & 3C48  &           &    & \multicolumn{2}{c|}{                           } & 16 / 14.6 & 62  & \multicolumn{2}{c}{ 9.3 $\times$ 5.0, $-$45.6} \\
1445$+$099 & 2008-07-22 & 3C147 &  6 /  5.1 & 52 & \multicolumn{2}{c|}{12.0 $\times$ 10.1,    84.1} & 32 / 31.2 & 62  & \multicolumn{2}{c}{ 5.2 $\times$ 4.2,    59.6} \\
1602$+$334 & 2009-06-06 & 3C286 &  6 /  5.1 & 51 & \multicolumn{2}{c|}{12.6 $\times$ 10.1,     5.8} & 32 / 31.4 & 51  & \multicolumn{2}{c}{ 5.7 $\times$ 5.0,     9.4} \\
1944$+$548 & 2013-07-12 & 3C48  & 16 / 15.3 & 57 & \multicolumn{2}{c|}{14.8 $\times$ 11.8, $-$50.8} & 32 / 31.0 & 57  & \multicolumn{2}{c}{ 6.1 $\times$ 5.1, $-$49.0} \\
2011$-$067 & 2008-12-06 & 3C286 &  6 /  4.9 & 25 & \multicolumn{2}{c|}{14.6 $\times$  9.6,    48.0} & 32 / 31.2 & 65  & \multicolumn{2}{c}{ 5.6 $\times$ 4.6,    48.1} \\
2212$+$018 & 2008-08-10 & 3C147 &  6 /  5.3 & 34 & \multicolumn{2}{c|}{14.8 $\times$ 11.0,    32.0} & 32 / 31.2 & 34  & \multicolumn{2}{c}{ 6.2 $\times$ 5.3,  $-$1.3} \\
2344$+$824 & 2013-07-24 & 3C48  & 16 / 15.1 & 49 & \multicolumn{2}{c|}{23.5 $\times$  9.4, $-$54.1} & 32 / 31.0 & 49  & \multicolumn{2}{c}{12.2 $\times$ 5.3, $-$62.4} \\
2355$+$498 & 2013-07-24 & 3C48  & 16 / 15.1 & 24 & \multicolumn{2}{c|}{15.0 $\times$  9.5, $-$52.1} & 32 / 31.0 & 50  & \multicolumn{2}{c}{ 8.5 $\times$ 4.7, $-$74.1} \\
\hline
 \multicolumn{11}{l}{Quasars} \\
0024$-$420 & 2013-07-05 & 3C48  & 16 / 15.2 & 67 & \multicolumn{2}{c|}{21.5 $\times$ 10.3,    21.6} & 32 / 31.2 & 84  & \multicolumn{2}{c}{10.8 $\times$ 6.6,    18.8} \\
0059$+$001 & 2006-07-22 & 3C48  &  6 /  5.2 & 36 & \multicolumn{2}{c|}{13.4 $\times$ 10.3,    57.3} & 32 / 31.3 & 67  & \multicolumn{2}{c}{ 6.3 $\times$ 4.7,    62.7} \\
0102$+$584 & 2013-07-12 & 3C48  & 16 / 15.3 & 55 & \multicolumn{2}{c|}{12.1 $\times$  9.6,  $-$6.9} & 32 / 31.3 & 57  & \multicolumn{2}{c}{ 5.9 $\times$ 4.6,     6.6} \\
0136$+$478 & 2013-07-12 & 3C48  & 16 / 15.3 & 37 & \multicolumn{2}{c|}{12.0 $\times$  9.7,    29.2} & 32 / 31.0 & 57  & \multicolumn{2}{c}{ 5.3 $\times$ 4.2,    27.2} \\
0217$+$738 & 2013-07-12 & 3C48  & 16 / 15.3 & 10 & \multicolumn{2}{c|}{18.6 $\times$  9.0,     1.8} & 32 / 31.0 & 58  & \multicolumn{2}{c}{ 7.7 $\times$ 4.5,    11.9} \\
0238$+$166 & 2007-08-17 & 3C48  &  6 /  5.4 & 45 & \multicolumn{2}{c|}{14.8 $\times$ 11.3, $-$89.5} & 32 / 30.8 & 45  & \multicolumn{2}{c}{ 7.4 $\times$ 4.7,    85.6} \\
0240$-$231 & 2008-07-21 & 3C147 &  6 /  5.1 & 63 & \multicolumn{2}{c|}{15.6 $\times$ 12.1, $-$17.7} & 32 / 31.2 & 63  & \multicolumn{2}{c}{ 6.9 $\times$ 5.2, $-$13.0} \\
0303$+$472 & 2009-10-25 & 3C48  &  6 /  5.0 & 99 & \multicolumn{2}{c|}{13.5 $\times$ 10.8,    66.9} & 32 / 31.4 & 99  & \multicolumn{2}{c}{ 6.2 $\times$ 5.3, $-$77.4} \\
0414$+$343 & 2013-07-26 & 3C48  & 16 / 15.1 & 57 & \multicolumn{2}{c|}{11.2 $\times$  7.9,    54.3} & 32 / 31.0 & 57  & \multicolumn{2}{c}{ 5.5 $\times$ 4.5,    73.1} \\
0555$+$398 & 2009-11-20 & 3C147 &           &    & \multicolumn{2}{c|}{                           } & 16 / 15.2 & 57  & \multicolumn{2}{c}{ 6.3 $\times$ 4.2,    75.2} \\
0607$-$085 & 2008-08-10 & 3C147 &  6 /  5.3 & 25 & \multicolumn{2}{c|}{15.2 $\times$ 11.2,    14.1} & 32 / 31.2 & 22  & \multicolumn{2}{c}{ 7.0 $\times$ 5.1,  $-$8.0} \\
0739$+$016 & 2008-07-21 & 3C147 &  6 /  5.1 & 61 & \multicolumn{2}{c|}{13.8 $\times$ 11.4, $-$75.4} & 32 / 31.2 & 61  & \multicolumn{2}{c}{ 6.3 $\times$ 4.8, $-$60.0} \\
1150$-$003 & 2008-07-22 & 3C48  &  6 /  5.1 & 67 & \multicolumn{2}{c|}{13.2 $\times$ 12.2,    71.1} & 32 / 31.2 & 67  & \multicolumn{2}{c}{ 5.7 $\times$ 4.7,    59.1} \\
1254$+$116 & 2008-07-21 & 3C147 &  6 /  5.1 & 54 & \multicolumn{2}{c|}{13.0 $\times$ 10.5,    63.1} & 32 / 31.2 & 54  & \multicolumn{2}{c}{ 5.1 $\times$ 3.6,    63.9} \\
1613$+$342 & 2002-12-21 & 3C286 &  6 /  4.9 & 48 & \multicolumn{2}{c|}{17.4 $\times$ 10.6, $-$79.7} &  6 /  5.4 & 48  & \multicolumn{2}{c}{ 5.2 $\times$ 4.9,    68.9} \\
1743$-$038 & 2006-07-20 & 3C286 &  6 /  5.2 & 55 & \multicolumn{2}{c|}{13.8 $\times$ 11.8,    56.6} & 32 / 31.3 & 54  & \multicolumn{2}{c}{ 6.8 $\times$ 4.9,    28.6} \\
1923$-$210 & 2008-07-22 & 3C147 &  6 /  5.1 & 53 & \multicolumn{2}{c|}{14.6 $\times$ 10.8,    23.3} & 32 / 31.2 & 53  & \multicolumn{2}{c}{ 6.4 $\times$ 4.3,    26.7} \\
2005$+$778 & 2013-07-12 & 3C48  & 16 / 15.3 & 56 & \multicolumn{2}{c|}{22.1 $\times$ 11.9, $-$37.7} & 32 / 31.0 & 57  & \multicolumn{2}{c}{ 8.6 $\times$ 4.6, $-$34.3} \\
2015$+$371 & 2013-08-13 & 3C48  & 16 / 15.0 & 57 & \multicolumn{2}{c|}{19.7 $\times$ 12.3,    84.1} & 32 / 31.0 & 57  & \multicolumn{2}{c}{ 8.1 $\times$ 5.0,    82.3} \\
2025$+$337 & 2013-08-13 & 3C48  & 16 / 15.0 & 55 & \multicolumn{2}{c|}{20.8 $\times$ 12.8,    79.2} & 32 / 31.0 & 55  & \multicolumn{2}{c}{ 8.5 $\times$ 5.0,    77.8} \\
2202$+$422 & 2013-08-02 & 3C48  & 16 / 15.0 & 48 & \multicolumn{2}{c|}{11.3 $\times$ 10.7,    65.2} & 32 / 31.0 & 48  & \multicolumn{2}{c}{ 7.0 $\times$ 4.9, $-$69.3} \\
2236$+$284 & 2008-08-10 & 3C147 &  6 /  5.3 & 21 & \multicolumn{2}{c|}{14.7 $\times$ 12.2, $-$44.7} & 32 / 31.2 & 21  & \multicolumn{2}{c}{ 6.9 $\times$ 5.0, $-$48.2} \\
2254$+$247 & 2013-07-24 & 3C48  & 16 / 15.1 & 44 & \multicolumn{2}{c|}{16.4 $\times$ 10.2,    75.5} & 32 / 31.0 & 44  & \multicolumn{2}{c}{ 9.9 $\times$ 4.7,    87.2} \\
\hline
 \multicolumn{11}{l}{Unidentified sources} \\
0110$+$565 & 2003-01-13 & 3C147 &  6 /  5.3 & 69 & \multicolumn{2}{c|}{14.8 $\times$ 11.0,    45.0} &  6 /  5.4 & 69  & \multicolumn{2}{c}{ 6.1 $\times$ 4.9,    54.5} \\
0321$+$123 & 2004-03-27 & 3C286 &           &    & \multicolumn{2}{c|}{                           } & 16 / 15.2 & 91  & \multicolumn{2}{c}{ 7.3 $\times$ 5.0,    17.4} \\
0438$+$488 & 2013-07-26 & 3C48  & 16 / 15.1 & 46 & \multicolumn{2}{c|}{13.0 $\times$  8.7,    53.5} & 32 / 31.0 & 57  & \multicolumn{2}{c}{ 6.0 $\times$ 4.6,    56.7} \\
0632$+$103 & 2004-07-24 & 3C147 &  6 /  4.9 & 63 & \multicolumn{2}{c|}{13.0 $\times$ 10.4,    71.6} & 32 / 31.0 & 64  & \multicolumn{2}{c}{ 7.7 $\times$ 4.1,    57.8} \\
1513$+$236 & 2008-07-27 & 3C286 &  6 /  5.1 & 44 & \multicolumn{2}{c|}{13.2 $\times$ 10.8,    72.0} & 32 / 31.2 & 49  & \multicolumn{2}{c}{ 5.6 $\times$ 4.5,    66.9} \\
2023$+$544 & 2013-07-12 & 3C48  & 16 / 15.3 & 32 & \multicolumn{2}{c|}{13.7 $\times$ 11.1, $-$23.4} & 32 / 31.0 & 56  & \multicolumn{2}{c}{ 6.0 $\times$ 5.3, $-$34.9} \\ [-0.3cm]
 & & & & & & & & & & \\
\hline
\end{tabular}
\label{obs-log}
\end{table*}

Typical GMRT continuum observations are
bracketed and interleaved by primary (flux density and bandpass) calibration source
scans on 3C\,48, 3C\,147 and 3C\,286 of 10--20 minute duration.
Interleaved in these observations are several
secondary (phase) calibration source scans, each of a few minutes long duration.
We extracted primary source and secondary calibration source scans from all the archival GMRT
235 MHz, 610 MHz data and dual frequency, both 235 MHz and 610 MHz data.
Table~\ref{obs-log} gives the details of the observations.
These continuum observations are usually performed in spectral line mode with
the total number of channels ranging from 64 to 256 and from 128 to 512 at
235 MHz and 610 MHz, respectively, and widths of each channel ranging from 32.5 kHz to 125 kHz.
Observations at 235 MHz for five sources, namely
0321$+$123, 0555$+$398, 0745$+$101, 0943$-$083 and 1400$+$621
were always more than 60\% affected by RFI or bad antennas 
and hence these results are not included here.

\subsection{Data Reduction}
\label{data-red}

\begin{table*}[ht]
\centering
\caption{The integrated flux density (in Jy) and the quality factor
for all 45 phase calibration sources.
Two qualities, good, `{\color{red}{G}}' or
moderate, `{\color{blue}{M}}' are provided based on
strength of the calibration source and strengths of contaminants (see Sect.~\ref{qual-fac}).
The typical error (1~$\sigma$) on flux density is $\lesssim$ 10~mJy.
The 235 MHz and 610 MHz are our GMRT measurements
and rest of the measurements are gleaned from the VLA \citet{vla-cal-manual},
which are from \citet{Kuhretal,Largeetal,Condonetal,Condonetal2,Griffithetal,Langstonetal,Beckeretal,WhiteBecker}. }
\begin{tabular}{l|rrrrrrrr|cc}
\hline
 & & & & & & & & & & \\ [-0.3cm]
 & \multicolumn{8}{c}{Frequency} & \multicolumn{2}{c}{Quality} \\
\cline{2-11} \\ [-0.3cm]
Calibrator & 235  & \multicolumn{1}{c|}{610} & 1.42 & 5.00 & 8.10 & 15.00 & 23.06 & 42.83 & \multicolumn{2}{c}{235~~610} \\
\multicolumn{1}{c|}{(J2000)} & \multicolumn{2}{c|}{(MHz)} &  \multicolumn{6}{c|}{(GHz)} & \multicolumn{2}{c}{(MHz)} \\
 & & & & & & & & & & \\ [-0.3cm]
\hline
 & & & & & & & & & & \\ [-0.3cm]
0010$-$418 & 3.57 & \multicolumn{1}{c|}{6.68} & 4.10 & 1.25 & 0.58 & 0.25 &      &      &{\color{blue}{M}} &{\color{red}{G}}  \\
0024$-$420 & 4.01 & \multicolumn{1}{c|}{3.23} & 2.80 & 1.70 & 0.94 & 0.40 &      &      &{\color{blue}{M}} &{\color{blue}{M}} \\
0029$+$349 & 1.02 & \multicolumn{1}{c|}{1.93} & 1.89 & 1.85 & 0.96 & 0.60 & 0.58 & 0.36 &{\color{blue}{M}} &{\color{blue}{M}} \\
0059$+$001 & 4.70 & \multicolumn{1}{c|}{3.61} & 2.50 & 1.35 & 0.96 & 0.70 &      & 0.42 &{\color{blue}{M}} &{\color{red}{G}}  \\
0102$+$584 & 0.88 & \multicolumn{1}{c|}{1.52} & 0.94 & 1.20 & 2.70 & 2.30 &      & 2.40 &{\color{blue}{M}} &{\color{blue}{M}} \\
0110$+$565 & 3.93 & \multicolumn{1}{c|}{2.78} & 1.90 & 0.85 & 0.53 & 0.30 &      & 0.11 &{\color{blue}{M}} &{\color{blue}{M}} \\
0119$+$321 & 4.15 & \multicolumn{1}{c|}{3.77} & 2.60 & 1.48 & 1.08 & 0.70 &      & 0.34 &{\color{blue}{M}} &{\color{red}{G}}  \\
0136$+$478 & 1.09 & \multicolumn{1}{c|}{1.24} & 1.62 & 1.88 & 1.80 & 1.60 &      & 1.60 &{\color{blue}{M}} &{\color{blue}{M}} \\
0217$+$738 & 0.93 & \multicolumn{1}{c|}{1.96} & 2.27 & 2.30 & 2.18 & 2.10 &      & 1.50 &{\color{blue}{M}} &{\color{blue}{M}} \\
0238$+$166 & 0.96 & \multicolumn{1}{c|}{1.49} & 1.26 & 1.73 & 1.30 & 3.30 &      & 3.50 &{\color{blue}{M}} &{\color{blue}{M}} \\
0240$-$231 & 2.16 & \multicolumn{1}{c|}{5.44} & 6.30 & 3.15 & 1.66 & 0.90 & 0.59 & 0.30 &{\color{blue}{M}} &{\color{red}{G}}  \\
0303$+$472 & 1.74 & \multicolumn{1}{c|}{0.94} & 1.80 & 2.47 & 1.43 & 2.90 &      & 1.20 &{\color{blue}{M}} &{\color{blue}{M}} \\
0321$+$123 &      & \multicolumn{1}{c|}{2.44} & 1.74 & 1.10 & 1.18 & 0.75 &      &      &                  &{\color{blue}{M}} \\
0410$+$769 & 9.65 & \multicolumn{1}{c|}{8.15} & 5.76 & 2.79 & 2.21 & 1.46 &      & 0.66 &{\color{blue}{M}} &{\color{red}{G}}  \\
0414$+$343 & 1.50 & \multicolumn{1}{c|}{1.97} & 2.03 & 1.50 & 1.23 & 0.97 &      & 0.50 &{\color{blue}{M}} &{\color{blue}{M}} \\
0438$+$488 & 3.95 & \multicolumn{1}{c|}{2.36} & 1.37 & 0.54 & 0.34 &      &      &      &{\color{blue}{M}} &{\color{red}{G}}  \\
0555$+$398 &      & \multicolumn{1}{c|}{0.49} & 1.70 & 5.00 & 6.20 & 2.80 &      & 2.50 &                  &{\color{blue}{M}} \\
0607$-$085 & 3.57 & \multicolumn{1}{c|}{3.02} & 3.00 & 2.70 & 3.22 & 2.10 &      & 3.60 &{\color{blue}{M}} &{\color{blue}{M}} \\ 
0632$+$103 & 3.74 & \multicolumn{1}{c|}{2.69} & 2.45 & 0.90 & 0.50 & 0.20 &      & 0.08 &{\color{blue}{M}} &{\color{red}{G}}  \\
0739$+$016 & 1.50 & \multicolumn{1}{c|}{1.64} & 1.95 & 1.80 & 2.00 & 2.05 &      & 2.10 &{\color{blue}{M}} &{\color{blue}{M}} \\
0745$+$101 &      & \multicolumn{1}{c|}{1.68} & 3.30 & 3.50 & 2.95 & 2.20 &      &      &                  &{\color{blue}{M}} \\
0943$-$083 &      & \multicolumn{1}{c|}{3.01} & 2.70 & 1.20 & 0.68 & 0.40 &      &      &                  &{\color{blue}{M}} \\
1130$-$148 & 4.87 & \multicolumn{1}{c|}{5.85} & 5.33 & 4.60 & 3.06 & 2.30 &      & 0.60 &{\color{blue}{M}} &{\color{red}{G}}  \\ 
1150$-$003 & 3.20 & \multicolumn{1}{c|}{3.64} & 2.80 & 1.92 & 1.25 & 1.40 &      & 0.65 &{\color{red}{G}}  &{\color{red}{G}} \\
1254$+$116 & 0.54 & \multicolumn{1}{c|}{0.85} & 1.00 & 0.80 & 0.90 & 0.90 &      & 0.70 &{\color{blue}{M}} &{\color{blue}{M}} \\ 
1313$+$675 & 6.39 & \multicolumn{1}{c|}{4.38} & 2.40 & 0.90 & 0.60 & 0.30 &      &      &{\color{blue}{M}} &{\color{red}{G}}  \\
1400$+$621 &      & \multicolumn{1}{c|}{6.96} & 4.40 & 1.72 & 1.08 & 0.67 & 0.48 & 0.28 &                  &{\color{red}{G}}  \\
1445$+$099 & 0.97 & \multicolumn{1}{c|}{2.39} & 2.60 & 1.20 & 0.73 & 0.40 &      & 0.10 &{\color{blue}{M}} &{\color{red}{G}}  \\
1513$+$236 & 2.14 & \multicolumn{1}{c|}{2.45} & 1.60 & 0.80 & 0.52 &      &      & 0.12 &{\color{blue}{M}} &{\color{red}{G}}  \\
1602$+$334 & 2.21 & \multicolumn{1}{c|}{3.03} & 2.60 & 2.00 & 2.05 & 1.40 &      & 0.41 &{\color{blue}{M}} &{\color{red}{G}}  \\
1613$+$342 & 2.29 & \multicolumn{1}{c|}{4.29} & 2.70 & 2.30 & 2.67 & 2.00 &      & 2.58 &{\color{blue}{M}} &{\color{red}{G}}  \\
1743$-$038 & 1.11 & \multicolumn{1}{c|}{1.29} & 1.55 & 2.70 & 3.80 & 3.80 &      & 5.10 &{\color{blue}{M}} &{\color{blue}{M}} \\
1923$-$210 & 1.18 & \multicolumn{1}{c|}{2.48} & 2.00 &      & 2.70 &      &      & 1.40 &{\color{blue}{M}} &{\color{red}{G}}  \\
1944$+$548 & 0.77 & \multicolumn{1}{c|}{1.96} & 1.66 &      & 0.67 &      &      & 0.15 &{\color{blue}{M}} &{\color{blue}{M}} \\
2005$+$778 & 0.87 & \multicolumn{1}{c|}{0.55} & 1.00 & 1.60 & 2.50 & 1.60 &      & 1.00 &{\color{blue}{M}} &{\color{blue}{M}} \\
2011$-$067 & 0.45 & \multicolumn{1}{c|}{2.24} & 2.60 & 1.30 & 0.85 & 0.49 &      & 0.80 &{\color{blue}{M}} &{\color{red}{G}}  \\
2015$+$371 & 1.07 & \multicolumn{1}{c|}{1.55} & 2.18 & 2.76 & 2.95 & 2.59 &      & 2.90 &{\color{blue}{M}} &{\color{blue}{M}} \\
2023$+$544 & 0.92 & \multicolumn{1}{c|}{1.33} & 1.20 & 1.05 & 1.00 & 1.10 &      & 0.70 &{\color{blue}{M}} &{\color{blue}{M}} \\
2025$+$337 & 0.92 & \multicolumn{1}{c|}{1.33} & 1.44 & 2.80 & 3.80 & 2.50 & 2.30 & 2.80 &{\color{blue}{M}} &{\color{blue}{M}} \\
2202$+$422 & 2.69 & \multicolumn{1}{c|}{4.78} & 6.07 & 5.40 & 3.95 & 3.50 &      & 2.50 &{\color{blue}{M}} &{\color{red}{G}}  \\
2212$+$018 & 5.35 & \multicolumn{1}{c|}{4.56} & 2.65 & 1.10 & 0.60 &      &      & 0.14 &{\color{blue}{M}} &{\color{red}{G}}  \\
2236$+$284 & 0.75 & \multicolumn{1}{c|}{0.83} & 1.50 & 2.00 & 0.90 & 1.50 &      & 0.80 &{\color{blue}{M}} &{\color{blue}{M}} \\
2254$+$247 & 3.29 & \multicolumn{1}{c|}{2.58} & 1.84 & 0.80 & 0.59 & 0.65 &      & 0.30 &{\color{red}{G}}  &{\color{red}{G}}  \\
2344$+$824 & 5.69 & \multicolumn{1}{c|}{5.46} & 3.79 & 1.34 & 0.75 & 0.40 &      & 0.30 &{\color{blue}{M}} &{\color{red}{G}}  \\
2355$+$498 & 1.77 & \multicolumn{1}{c|}{2.74} & 2.36 & 1.60 & 1.00 & 0.90 &      & 0.28 &{\color{blue}{M}} &{\color{blue}{M}} \\ [-0.3cm]
 & & & & & & & & & & \\
\hline
\end{tabular}
\label{cal-data}
\end{table*}

The raw telescope format data were converted to FITS and then analysed in \textsc{aips}
using standard procedures.
The flux density calibrators were used as an amplitude calibrator and to correct the bandpass shape.
We used an extension of the \citet{Baarsetal} scale to low frequencies
and the uncertainty, both due to calibration and systematic typically is $\lesssim$~5\%
\citep[see also][for a detailed discussion on the error in the estimated
flux densities]{lalandrao,lalandrao05}.
The data were carefully inspected for bad antennas, scintillations and intermittent RFI,
which were all flagged.
Occasionally, we also used flagging and calibration (\textsc{flagcal}) software pipeline
\citep{prasadchengalur} for automatic flagging and calibration of data.
Typically $\lesssim$20\% data were flagged for each calibration source.
We left 3-5 channels on either side of the bandpass and
the central channels of this flagged and calibrated data were averaged using the
task \textsc{splat} to reduce data volume by a factor of 10--100 and
making sure that the final model images will not be affected 
by bandwidth smearing \citep{cotton1999,bridleschwab1999}.

The averaged data were subsequently used to make continuum images using
the \textsc{aips} task \textsc{imagr}.
While imaging, a mosaic of 55 and 37 slightly overlapping facets
covering fields-of-view of $\sim$3.2 deg$^2$ and $\sim$0.5 deg$^2$
were used at 235 MHz and 610 MHz, respectively, to account for the
non-coplanarity of the incoming wavefront \citep{cotton1999,perley1999}.
Throughout the analysis, we ran \textsc{imagr} in 3--D mode for \textit{w}--term
correction and used `uniform' weighting \citep{briggsetal1999}.
The presence of a large number of sources in the field-of-view in each mapped field
allowed us to perform 2--3 rounds of phase-only self-calibration \citep{PR1984,CF1999},
which was sufficient for the self-calibration process to converge.
At each round of self-calibration, the image and the visibilities were compared to check for the improvement of the source model.
While imaging, we performed deep \textsc{clean}ing
so that the background is thermal noise-dominated and no depression/bowl
is seen, and therefore we do not make any deconvolution errors.
The final mosaic of facet-images were stitched using task \textsc{flatn} to
construct a final model-image and corrected for the primary beam of the GMRT antennas.
Below, in Sect.~\ref{results},
we present results from these radio images at default angular resolutions
(see Table~\ref{obs-log}).

\section{Results}
\label{results}

\subsection{Quality Factor}
\label{qual-fac}

There are several criteria for choosing and including a calibration source,
e.g., for correcting instrumental, geometric and atmospheric gains,
both amplitude and phase variations,
monitoring the quality of the data and for searching
for occasional amplitude and phase jumps, if any.
Unlike the quality standards, e.g., `P', `S', etc. of phase calibration sources from
the VLA \citet{vla-cal-manual} for the desired observing frequency and array configuration,
which depend on morphology of each calibration sources
and hence ($u,v$)$_{\rm min}$ and ($u,v$)$_{\rm max}$, it is difficult to
provide similar quality classifications for GMRT calibration sources. 
This is because GMRT has a hybrid configuration (see also Sect.~\ref{sample})
whereby it samples the ($u,v$) plane adequately on the short baselines
as well as on the long baselines, and
its data often suffers from broadband RFI,
and hence quantifying in absolute units of closure errors and therefore
ascribing ($u,v$)$_{\rm min}$ and ($u,v$)$_{\rm max}$ limits to a calibration source
is inefficient.
Additionally, all models of calibration sources are self-calibrated,
which provide us with adequate image models.

Here, instead we provide two qualities, good, `{\color{red}{G}}' or
moderate, `{\color{blue}{M}}' based on
strength of the calibration source and strengths of all other sources detected
in the field-of-view of the calibration source, which are above five times the root-mean-square noise.
We henceforth describe all other sources detected
in the calibration source field as the contaminants.
Quantitatively, if a phase calibration source at the phase-centre has
flux density $\gtrsim$~1~Jy and the contaminants in its field-of-view contribute
up to $\lesssim$0.01~Jy is called a `good' quality calibration source.
The rest of the sources are qualified as `moderate' in this criteria at both
frequencies.
Table~\ref{cal-data} lists good, {\color{red}{G}} or moderate, {\color{blue}{M}}
for each of the 45 low frequency calibration sources at 235 MHz and 610 MHz.
To compare this criterion of good or moderate with respect to criteria of
the VLA \citet{vla-cal-manual}, the contribution to the expected
closure errors due to unmodeled phase of the contaminants
corresponds to 10\% or less for good, and
more than 10\% for moderate calibration sources.
In other words, our criteria suggests that a good, {\color{red}{G}}
phase calibration source is either of the P-class or the S-class (see also Sect.~\ref{sample}),
and a moderate, {\color{blue}{M}} phase calibration source is
confused or unknown or inappropriate according to the quality
defined by the VLA \citet{vla-cal-manual}.
Note that since, we are providing a model for each calibration source field,
i.e., deconvolved restored maps, the closure error is much less, $\lesssim$3\%,
irrespective of whether the calibration source is a good, {\color{red}{G}}
or a moderate, {\color{blue}{M}} quality.

\begin{figure*}
\begin{center}
\begin{tabular}{cc}
\includegraphics[angle=270,width=8.3cm]{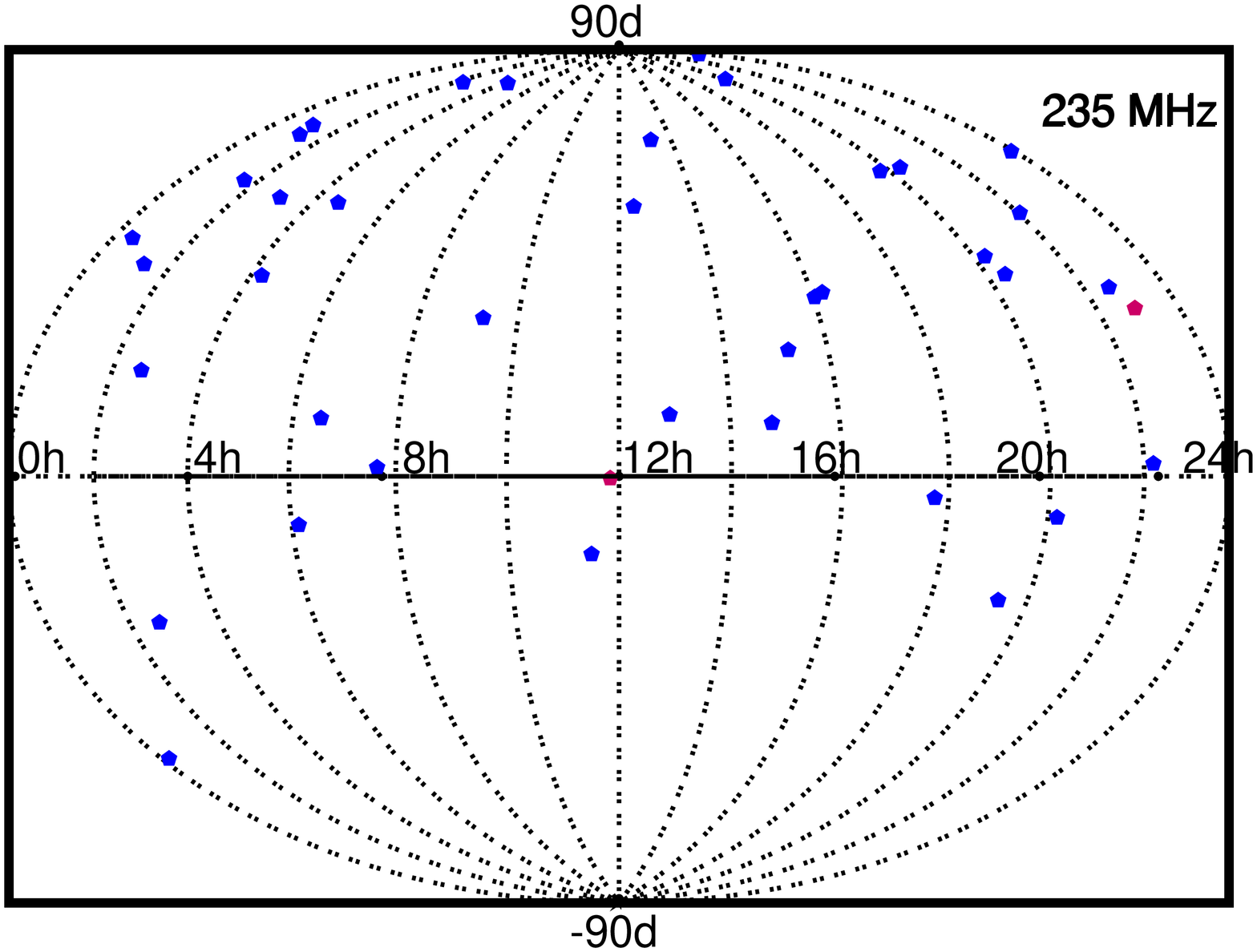} &
\includegraphics[angle=270,width=8.3cm]{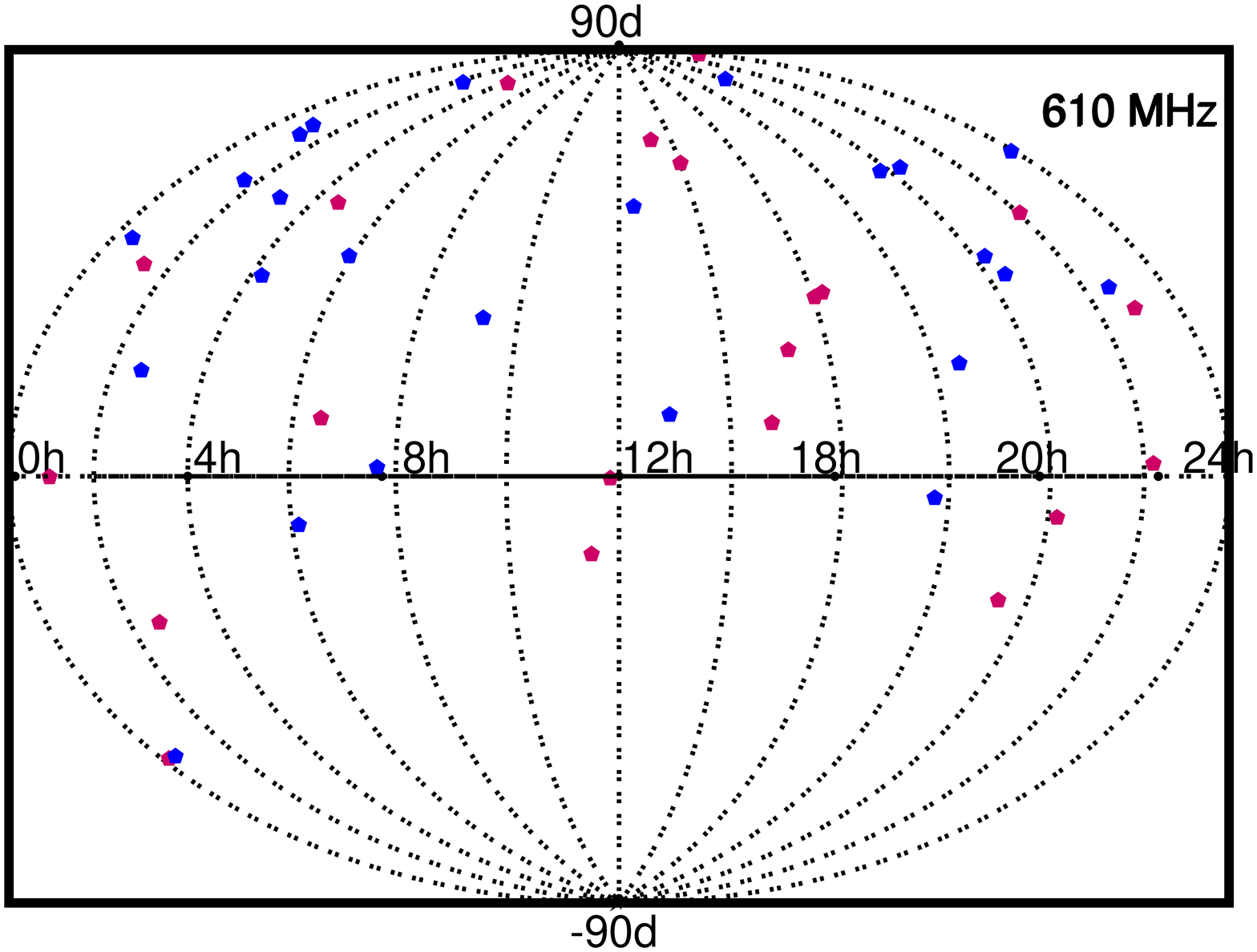}
\end{tabular}
\caption{A Mollweide projection plot in equatorial coordinates showing
the sky distribution of sources at 235 MHz (left-hand panel) and
at 610 MHz (right-hand panel).
The horizontal and vertical axes in each panel are right ascension and
declination, respectively.
The legends in the plots, i.e., red and blue points, correspond to
the quality factor, good, `{\color{red}{G}}' and moderate, `{\color{blue}{M}}', respectively
(see Sect.~\ref{qual-fac}) of the calibration source.}
\label{fig2}
\end{center}
\end{figure*}

\subsection{Sky Distribution} \label{sky_dist}

The distribution of these phase calibration radio sources on the sky reflects
the structure of the observable universe on the largest possible scales that
can be mapped with the GMRT via the phase-referencing technique.
Fig.~\ref{fig2} shows the distribution of phase calibration sources on the
celestial sphere.
The smallest and the largest angular separations between any two phase calibration
sources are $\sim$0.5~deg and $\sim$52~deg, respectively.
Thus, it is possible that for a certain observation the target source
may be as far as $\sim$26~deg from the nearest phase calibration source.
Although Fig.~\ref{fig2} suggests the distribution of phase calibration sources looks
patchy, there are no visible large voids in the sky.
In order to provide support to this suggestion,
we test the hypothesis that these phase calibration
sources are randomly distributed in the sky,
and hence the right ascension and sin($\pi/2 - {\rm declination}$) plane will be
uniformly  distributed.
We constructed a Halton sequence of size equal to the number of calibration
sources. The Halton sequence is a deterministic sequence that
produces well-spaced, uniformly distributed data,
and we call this uniformly distributed sequence over the right ascension
and sin($\pi/2 - {\rm declination}$) plane the null distribution.
We compute the gap statistics \citep{Tibshiranietal},
a method for estimating the number of clusters
or its nature of distribution in a set of data.  Mathematically, let us consider
$n$ uniform data points in $p$ dimensions, with $k$ centres, which
align themselves in an equally spaced manner; then
$$
\Sigma~{\rm log}(W_k) \simeq {\rm log}(pn/12) - (2/p)~{\rm log}(k) + {\rm constant},
$$
where $\Sigma~{\rm log}(W_k)$ is the expectation of the error measure, log~$(W_k)$.
Normalizing the graph of log~$(W_k)$ by comparing it with its expectation
under the null distribution of the data gives a quantitative estimate
of the uniform nature of observed distribution of phase calibration sources.
We compared the gap statistics from observed and expected distributions of data
and found the observed and expected curve are very close
and the gap estimate is unity, suggesting uniform distribution of phase calibration sources.
Alternatively, the two-point correlation function 
is also commonly used to quantify the clustering of distribution in a set of data.
\citet{DavisPeebles1983} defined an estimator
$$
\rho = [(\eta_R / \eta_D) \times (DD / DR)] - 1
$$
where $DD$ and $DR$ are counts of pairs of sources (in bins of separation)
in the set of observed data distribution and
between the set of observed data and set of null data distributions,
and $\eta_D$ and $\eta_R$ are the mean number densities of sources in the observed data
and null data distributions, respectively.
We find $< \rho > \simeq -0.008$ between the observed distribution of phase
calibration sources and the null distribution constructed using the Halton
sequence, again suggesting
uniform distribution of phase calibration sources, i.e.,
the right ascension and sin($\pi/2 - {\rm declination}$) plane is
uniformly distributed.

Recall that the observations of phase calibration sources are important for
(i) tracking the instrumental and the atmospheric gains,
(ii) monitoring the quality and sensitivity of the data
and (iii) spotting the occasional amplitude and phase jumps.
Hence one chooses the calibration source closest to the target source to better
calibrate atmospheric gain fluctuations.
The sky distribution discussed here suggests that the phase calibration sources are
uniformly distributed, but presently when making a choice with this database of
calibration sources, the calibration source and
the target source may have relatively large, $>$~20~deg angular separation
than the recommended proximity between the calibration source and the target source
\citep{vla-status};
clearly, our ongoing efforts to increase the size of the database would
address this concern.

\subsection{Comparison to a Conventional Image}
\label{comparison}

\begin{figure*}
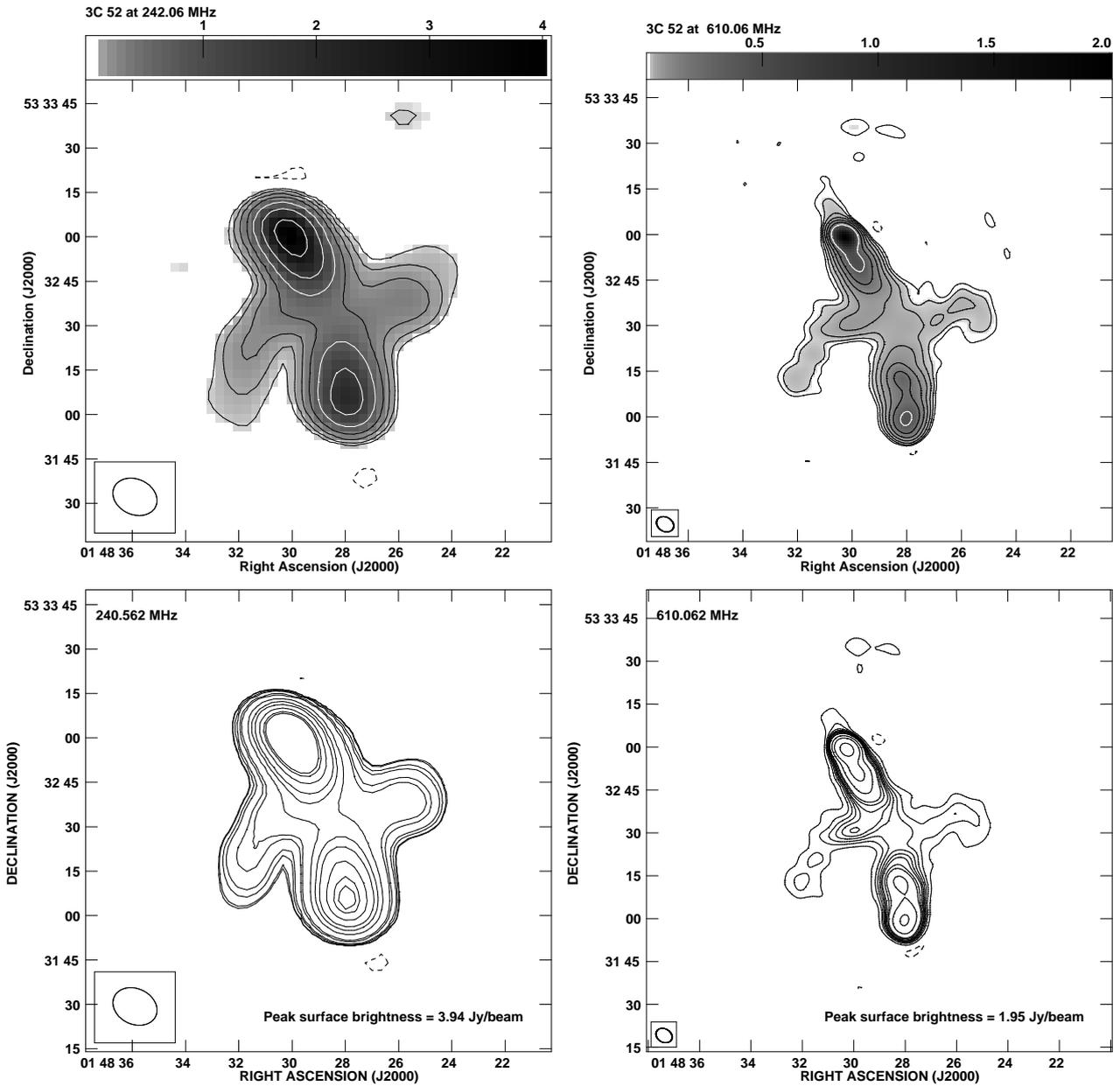

\begin{center}
\begin{tabular}{cc}
\includegraphics[width=8.6cm]{fig2a.PS} &
\hspace*{-0.7cm} \includegraphics[width=8.6cm]{fig2b.PS} \\ [-1.2cm]
\includegraphics[width=8.6cm]{fig2c.PS} &
\hspace*{-0.7cm} \includegraphics[width=8.6cm]{fig2d.PS} \\ [-0.8cm]
\end{tabular}
\caption{Upper panel: full synthesis GMRT maps of 3C\,52 at 235 MHz (left-hand panel)
and 610 MHz (right-hand panel).
The data are reduced using the image model from the database 
to obtain the calibration for the secondary phase
calibration source; the data reduction recipe is explained in Sect.~\ref{comparison}.
The synthesized beams for 235 MHz and 610 MHz are 15$^{\prime\prime}$.33 $\times$ 11$^{\prime\prime}$.92 at a P.A. of 64$^{\circ}$.46
and
5$^{\prime\prime}$.92 $\times$ 4$^{\prime\prime}$.68 at a P.A. of 58$^{\circ}$.35, respectively.
The peak surface brightness and the RMS noises in the immediate vicinity of the source for 235 MHz and 610 MHz maps are
4.0 Jy~beam$^{-1}$ and 2.0 Jy~beam$^{-1}$,
and 4.9 mJy~beam$^{-1}$ and 0.6 mJy~beam$^{-1}$, respectively.
The contour levels in these two maps are
RMS noise level $\times$ $-$3, 3, 6, 12, 24, 48, 96, 192 mJy~beam$^{-1}$.
Lower panel: the original published, full synthesis GMRT maps of 3C 52 at 235 MHz
(left-hand panel) and 610~MHz (right-hand panel),
Fig~2, upper-panel images of \citet{lalandrao}.
The synthesized beams for 235 MHz and 610 MHz maps are
15$^{\prime\prime}$.4~$\times$~12$^{\prime\prime}$.0 at a P.A. of 63$^{\circ}$.7
and
5$^{\prime\prime}$.9~$\times$~4$^{\prime\prime}$.7 at a P.A. of 57$^{\circ}$.5,
respectively; and the contour levels in the two maps are, respectively,
$-$50, 50, 60, 80, 100, 160, 200, 400, 600, 800, 1200, 1600, 1800 mJy~beam$^{-1}$
and
$-$8, 8, 20, 30, 40, 50, 60, 80, 100, 200 mJy~beam$^{-1}$,
The error-bars in the full synthesis maps found at a source free location
are $\sim$1.4~mJy~beam$^{-1}$ and $\sim$0.3~mJy~beam$^{-1}$ at 235 and 610~MHz, respectively. }
\label{3c52-im}
\end{center}
\end{figure*}

In order to test our methodology, we performed data reduction in \textsc{aips}
using standard procedures (see Sect.~\ref{data-red}),
but using the appropriate model image file from the database.
We typically run following steps during the standard data reduction procedures, namely,
\begin{itemize}
\item[(1)] loading the FITS file (task \textsc{fitld}),
\item[(2)] entering the absolute flux density of the primary calibration source (task \textsc{setjy}),
\item[(3)] determining calibration for the scans of primary flux density and secondary phase
calibration sources (task \textsc{calib}),
\item[(4)] bootstrapping the flux density (task \textsc{getjy}),
\item[(5)] interpolating the gains (task \textsc{clcal}),
\item[(6)] perform bandpass calibration (task \textsc{bpass}), and
\item[(7)] making image of the target source (task \textsc{imagr}).
\end{itemize}
Instead, we used the model image file from the database
to determine calibration for primary and secondary calibration sources,
hence, step~(3) mentioned above, was performed via. following two steps,
\begin{itemize}
\item[(3a)] solve for the antenna-based gains for primary flux density calibration source, and
\item[(3b)] use the image model from the database, either the \textsc{clean}-component file or
the model image file, to obtain this calibration for the secondary phase
calibration source, i.e., $\phi_{\rm true}^{\rm c}$;
\end{itemize}
and rest of the data reduction steps were performed in the usual standard manner.

As a case study,
we made an image of the source 3C\,52 (project code 03DVL01, observing date
12 January 2003) using a image model from the database for 0110$+$565 phase
calibration source of moderate {\color{blue}{M}} quality (see
Table~\ref{cal-data}) and compared it with the image published in
\citet{lalandrao}. The upper-panel of
Fig.~\ref{3c52-im} shows the radio images at 235 MHz and 610 MHz
using image model of phase-calibration source 0110$+$565 from the database,
while the lower-panel shows the original published image \citep[Fig~2:][]{lalandrao}.
Clearly, we reproduce all morphological details at both these frequencies.
To make further comparisons of the two images, we made difference images,
subtracting upper-panel images from the corresponding lower-panel original
published images (Fig.~\ref{3c52-im}),
at both these frequencies after matching their angular resolutions,
and the RMS difference in the resultant maps was less than 4\% and 3\% at
235 MHz and 610 MHz, respectively.
We performed similar analysis on several radio galaxies published in earlier papers
e.g., \citet{lalandrao,laletal} 
and the RMS difference in the resultant maps was always $\lesssim$4\%.
We conclude that the methodology is robust and does not suffer from any systematics;
thereby providing us confidence in using the image model from the database to perform
phase calibration and thus fully exploiting the technique of phase-referencing.

Presently, for pulsar observations,
the calibration source is used as a point source for phasing the array
using an observatory based tool,
{\sc rantsol}.  {\sc rantsol}
processes the raw telescope format data in real-time
and determines the calibration to be applied on to the data
given a model of the calibration source in order to phase the array.
The output data stream would have these corrections applied.
Hence only a handful of bright, 
P-class or S-class VLA calibration sources with
flux densities more than a few Jansky are used at the observatory.
Instead, a suitable image model from the GMRT calibrator database can be used to
obtain antenna based solutions for all antennas.
These antenna solutions would thus allow the phasing of the GMRT array for further
processing by the pulsar receiver system.

\begin{figure*}
\begin{center}
\begin{tabular}{c}
\includegraphics[width=15.8cm]{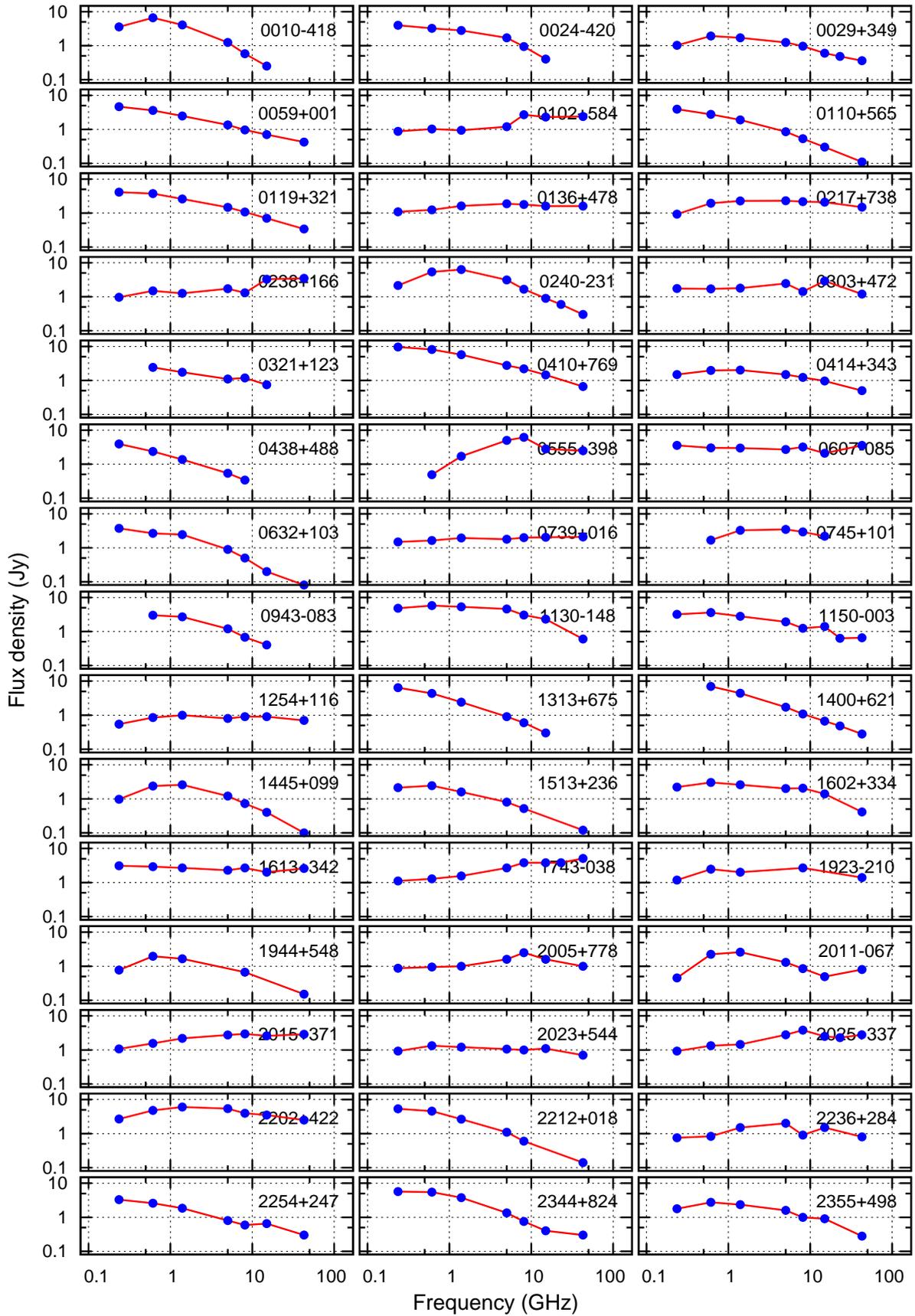} \\ [-0.4cm]
\end{tabular}
\caption{Integrated flux density (spectra) for the calibration sources.
Various measurements along with error bars (not plotted) at 235 MHz and 610 MHz are
explained in Table~\ref{cal-data}.}
\label{integ-235-610}
\end{center}
\end{figure*}

\subsection{Integrated Radio Spectra}
\label{int-rad-spec}

\begin{table*}
\caption{Shapes of spectra for the three categories of phase calibration sources.}
\begin{tabular}{l|lll}
\hline
 & & & \\ [-0.3cm]
 & \multicolumn{1}{c}{Straight} & \multicolumn{1}{c}{Curved} & \multicolumn{1}{c}{Complex} \\
 & & & \\ [-0.3cm]
\hline
 & & & \\ [-0.3cm]
Galaxies     & 0119$+$321, 0410$+$769, 0943$-$083, & 0010$-$418, 0029$+$349, 0745$+$101, & 2011$-$067, 2344$+$824, 2355$+$498 \\
             & 1313$+$675, 1400$+$621, 2212$+$018  & 1130$-$148, 1445$+$099, 1602$+$334, & \\
             &                                     & 1944$+$548                          &  \\
Quasars      & 0059$+$001, 0136$+$478, 0607$-$085, & 0024$-$420, 0217$+$738, 0240$-$231, & 0102$+$584, 0238$+$166, 0303$+$472, \\
             & 0739$+$016, 1613$+$342, 1743$-$038, & 0414$+$343, 2202$+$422              & 0555$+$398, 1150$-$003, 1254$+$116, \\
             & 2015$+$371                          &                                     & 1923$-$210, 2005$+$778, 2025$+$337, \\
             &                                     &                                     & 2236$+$284, 2254$+$247  \\
Unidentified & 0110$+$565, 0321$+$123, 0438$+$488, & 1513$+$236    & 0632$+$103 \\
             & 2023$+$544                          &               & \\
\hline
\end{tabular}
\label{spectra-shape}
\end{table*}

\begin{figure*}
\begin{center}
\begin{tabular}{ccc}
\includegraphics[width=4.1cm,angle=-90]{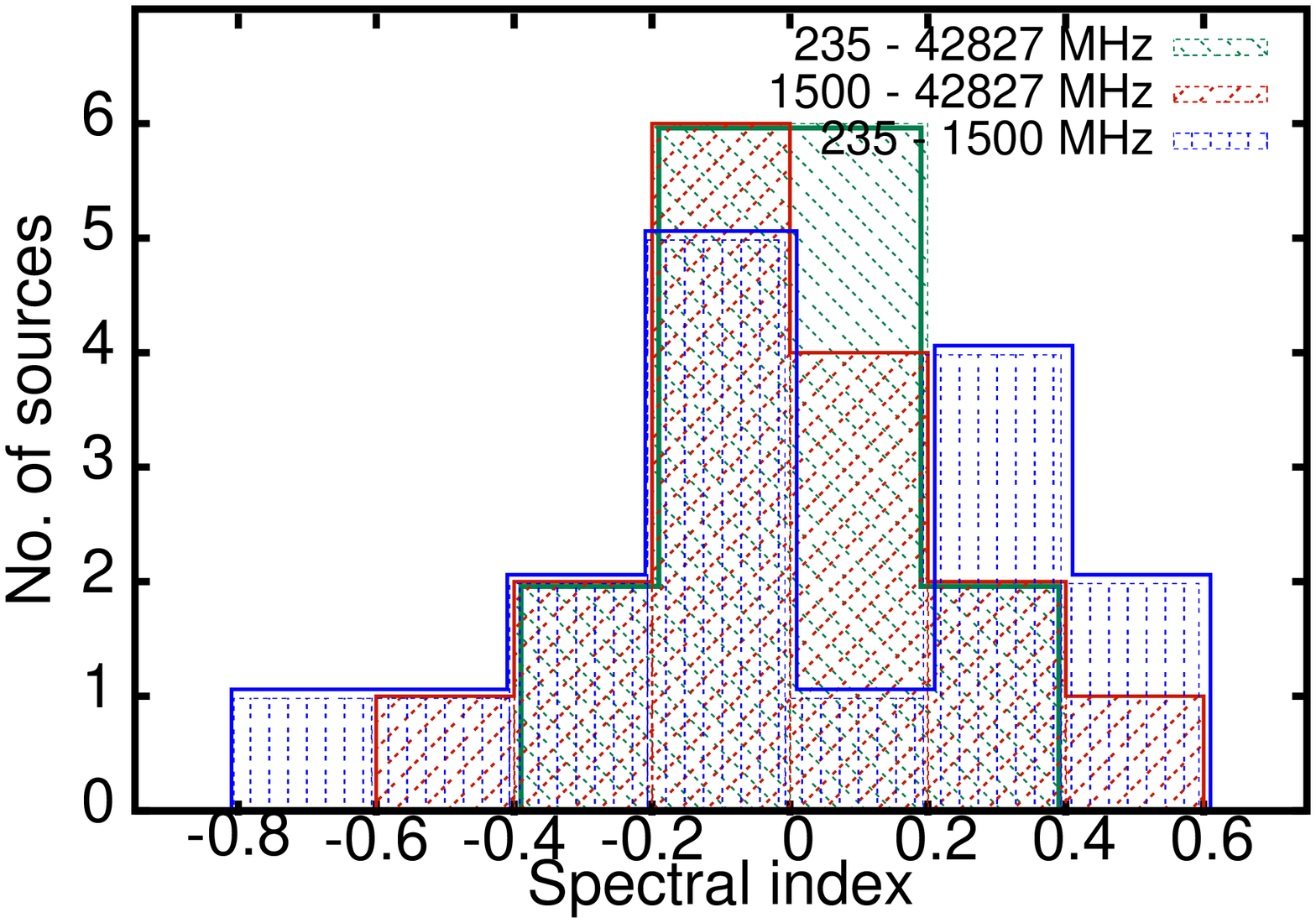} &
\hspace*{-0.3cm} \includegraphics[width=4.1cm,angle=-90]{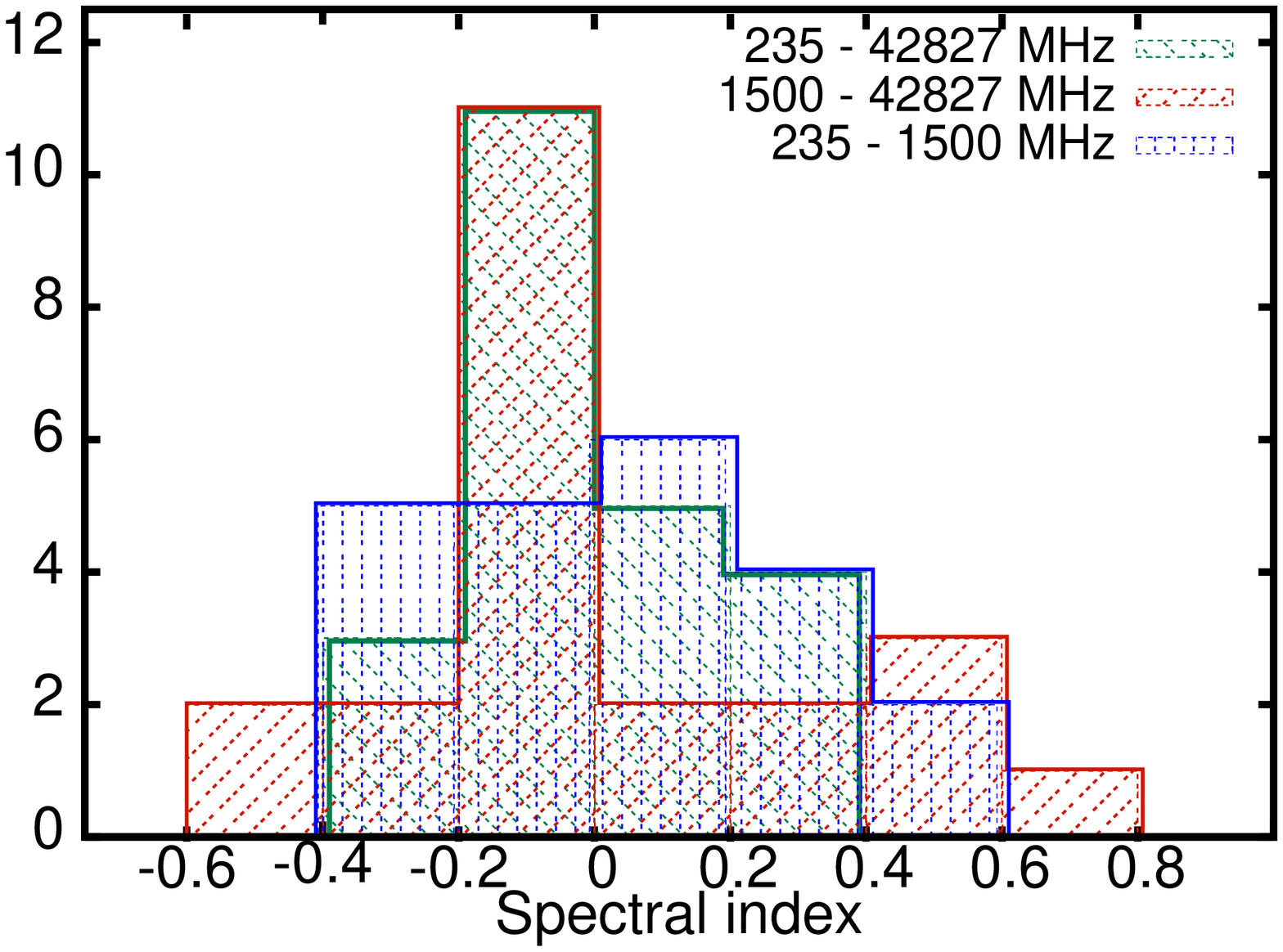} &
\hspace*{-0.5cm} \includegraphics[width=4.1cm,angle=-90]{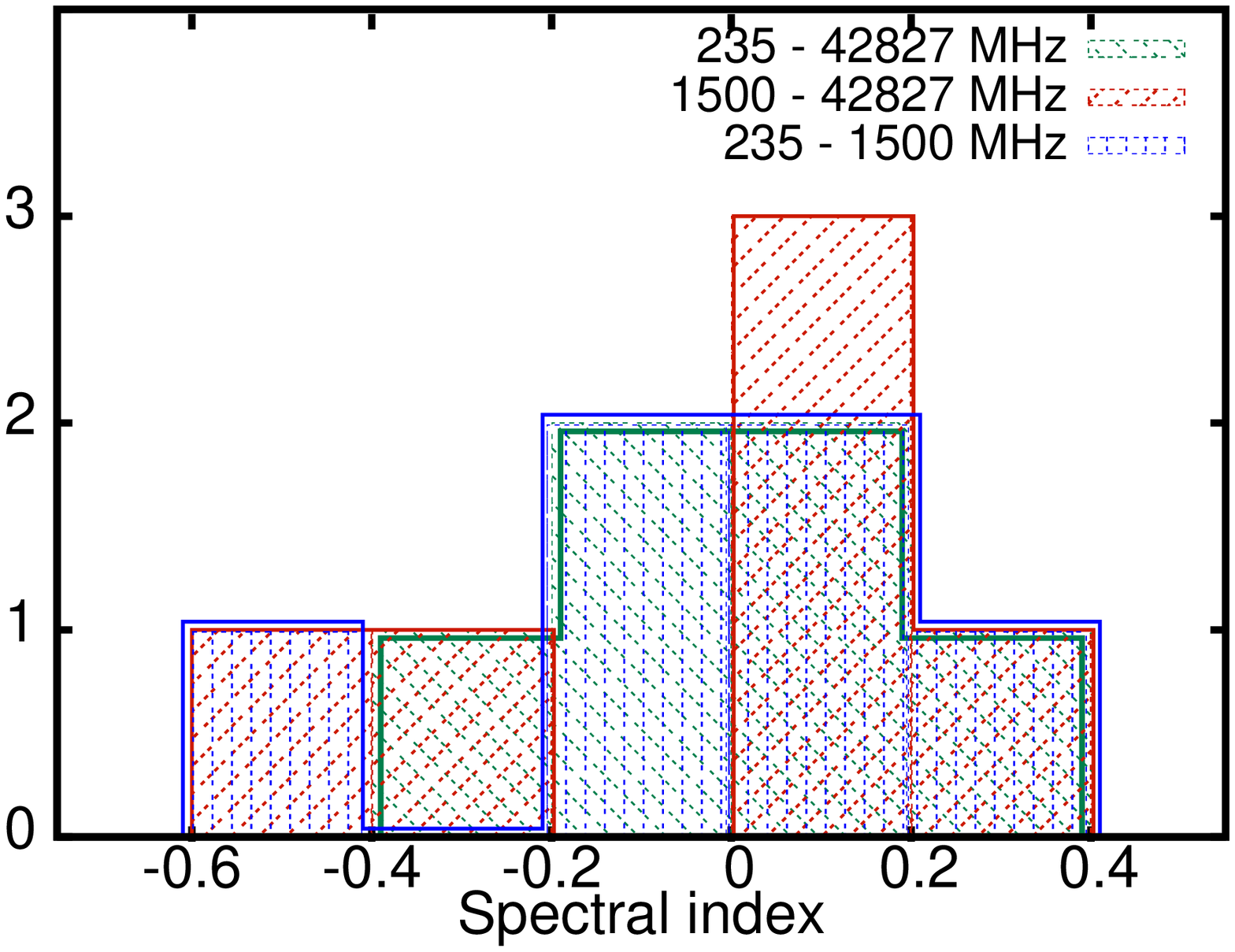}
\end{tabular}
\caption{Histograms showing the distribution of spectral indices for calibration sources
(a) at low frequencies, $\lesssim$~1500~MHz (blue), (b) at high frequencies $\gtrsim$~1500~MHz
(red), and (c) over a complete frequency range from 235 MHz to 42.83 GHz (green)
for all 45 phase calibration sources.
(The borders are shifted to avoid overlapping lines and hence, for clarity.)
Three panels correspond to radio galaxies (left-hand panel), quasars (middle-panel)
and the unidentified sources (right-hand panel).}
\label{hist-alpha}
\end{center}
\end{figure*}

Determining the radio spectra of phase calibration sources is
important to understand their nature and more importantly flux density scales.
We analysed the integrated radio spectra of phase calibration sources by
complementing the GMRT measurements at 235 MHz and 610 MHz with flux densities
at other frequencies available in literature.
Since all the sample phase calibration sources are
unresolved; the flux densities have been determined using \textsc{aips} task \textsc{jmfit}.
The integrated spectra are reported in Table~\ref{cal-data}.
In Fig.~\ref{integ-235-610}, we present the plots of the integrated radio
spectra along with a power law fit to the data.
To a large extent, the GMRT measurements agree both with the fit and with
the adjacent data points taken from the VLA \citet{vla-cal-manual} database,
thus providing confidence in the flux density scale in our images.

It is clear that there is considerable variety in the spectral shapes.
Based on these spectral shapes \citep{keller-paulini-will}, the spectra are
qualitatively classified as:
(i) straight, (ii) curved, both concave or convex and (iii) complex.
These categories are the same as those defined by \citet{LP1980,keller-paulini-will,Kuhretal} in their
study of radio loud quasars and radio galaxies.
Briefly,
sources showing straight spectra have a simple power law over the whole frequency range,
sources showing curved, concave or convex spectra have positive or negative curvature, respectively,
and
sources showing complex spectra are thought to be the result of
two or more components with different spectra.
Table~\ref{spectra-shape} summarizes various classes of radio spectra
for these calibration sources
and their radio spectra are illustrated in Fig.~\ref{integ-235-610}.
For the straight radio spectra, the slopes tend to be steep,
in the range $-1.0 < \alpha < -0.5$, where $S_\nu \propto \nu^\alpha$,
and $S_\nu$ and $\nu$ are flux density and frequency, respectively.
Whenever the different components of a curved or a complex source have
significantly different spectral indices, the net spectrum shows
positive curvature.  This suggests that when a source is made up of
several components, the spectral indices and hence the electron-energy
distribution of the various components are usually similar.
Three sources, 1150$-$003, 1923$-$210 and 2005$+$778, all quasars,
the spectra are extremely complex with several maxima and minima.
Such spectra are shown by variable radio sources \citep{keller-paulini}
and are probably the sum of a number of components each of which has a
spectral cutoff at a different frequency.

Since there is no clear simple form that can be used to represent all radio spectra,
we therefore fit the spectrum of each source over two ranges of frequency
and a complete range of frequency.
Fig.~\ref{hist-alpha} shows
histograms of the distribution of spectral indices for three categories,
radio galaxies, quasars and the unidentified, of calibration sources
at low frequencies $\lesssim$~1500~MHz, high frequencies $\gtrsim$~1500~MHz
and over a complete frequency range from 235 MHz to 42.83 GHz.
We also provide mean spectral index and dispersion for the groups of sources in
Table~\ref{mean-alpha} for these ranges of frequency.
A summary of shapes of radio spectra
and their associations with the categories of sources are as follows:
\begin{itemize}
\item[$-$] The spectra are about equally divided between straight, curved
(either concave or convex) and complex shapes.
\item[$-$] Quasars tend to exhibit flatter radio spectra as compared to the
radio galaxies or the unidentified sources, both at low ($\lesssim$ 1500 MHz) and
at high ($\gtrsim$ 1500 MHz) frequencies (see Table~\ref{mean-alpha}).
This difference is possible because quasars have small, $<$~1~arcsec angular diameters
and in these sources synchrotron self-absorption has made a major contribution
to the flattening of the spectrum at low frequencies.
\item[$-$] Quasars are known to be radio variable and hence possibly show complex spectra
more frequently.
\item[$-$] Radio galaxies are found to have systematically
steeper spectra than either the quasars or the unidentified sources.
The steeper spectra are possibly either due to the spectral ageing of
synchrotron-emitting radio lobes or due to the large redshift of distant galaxies
causing the shift of the spectrum to lower frequencies.
\end{itemize}

\begin{table*}
\centering
\caption{Mean spectral index and dispersion for the groups of sources.}
\begin{tabular}{l|ccc}
\hline
 & & & \\ [-0.3cm]
 & 240--1500 MHz & 1500--42827 MHz & 240--42875 MHz \\
 & & & \\ [-0.3cm]
\hline
 & & & \\ [-0.3cm]
Galaxies     & $< \alpha > = ~0.03 \pm0.18$ & $< \alpha > = -0.61 \pm0.06$ & $< \alpha > = -0.31 \pm0.10$  \\
Quasars      & $< \alpha > = ~0.19 \pm0.09$ & $< \alpha > = -0.13 \pm0.09$ & $< \alpha > = -0.02 \pm0.07$ \\
Unidentified & $< \alpha > = -0.27 \pm0.09$ & $< \alpha > = -0.57 \pm0.05$ & $< \alpha > = -0.38 \pm0.10$ \\
\hline
\end{tabular}
\label{mean-alpha}
\end{table*}

Note that there is evidence for time variability in the radio emission from
some calibration sources \citep{Barvainisetal},
in particular the quasars, which are known to show radio variability,
one must be cautious in interpreting non-simultaneous radio flux density data.
Furthermore, it is unlikely that the time variability is an issue at GMRT frequencies
because it is the compact core emission that shows time variability, but
the low frequency emission is dominated by extended, low-surface brightness,
diffuse emission, which has steeper spectra and does not show time variability
\citep{Lal2009,Lal15,LP1980}.
In addition, the LOFAR Multifrequency Snapshot Sky Survey \citep[MSSS,][]{Healdetal} and
the GaLactic and Extragalactic All-Sky MWA Survey \citep[GLEAM,][]{Waythetal}
would constrain the spectral properties of these calibration sources. 
However, the motivation of the paper is to build a database of low frequency
GMRT calibrator manual, and therefore the presence or absence of time variability
is not an issue as long as models for calibration sources do not change,
since we use these models to determine the phase calibration for the secondary
phase calibration sources.

\section{Summary}
\label{summary}

\begin{figure*}
\begin{center}
\begin{tabular}{c}
\includegraphics[width=17.0cm]{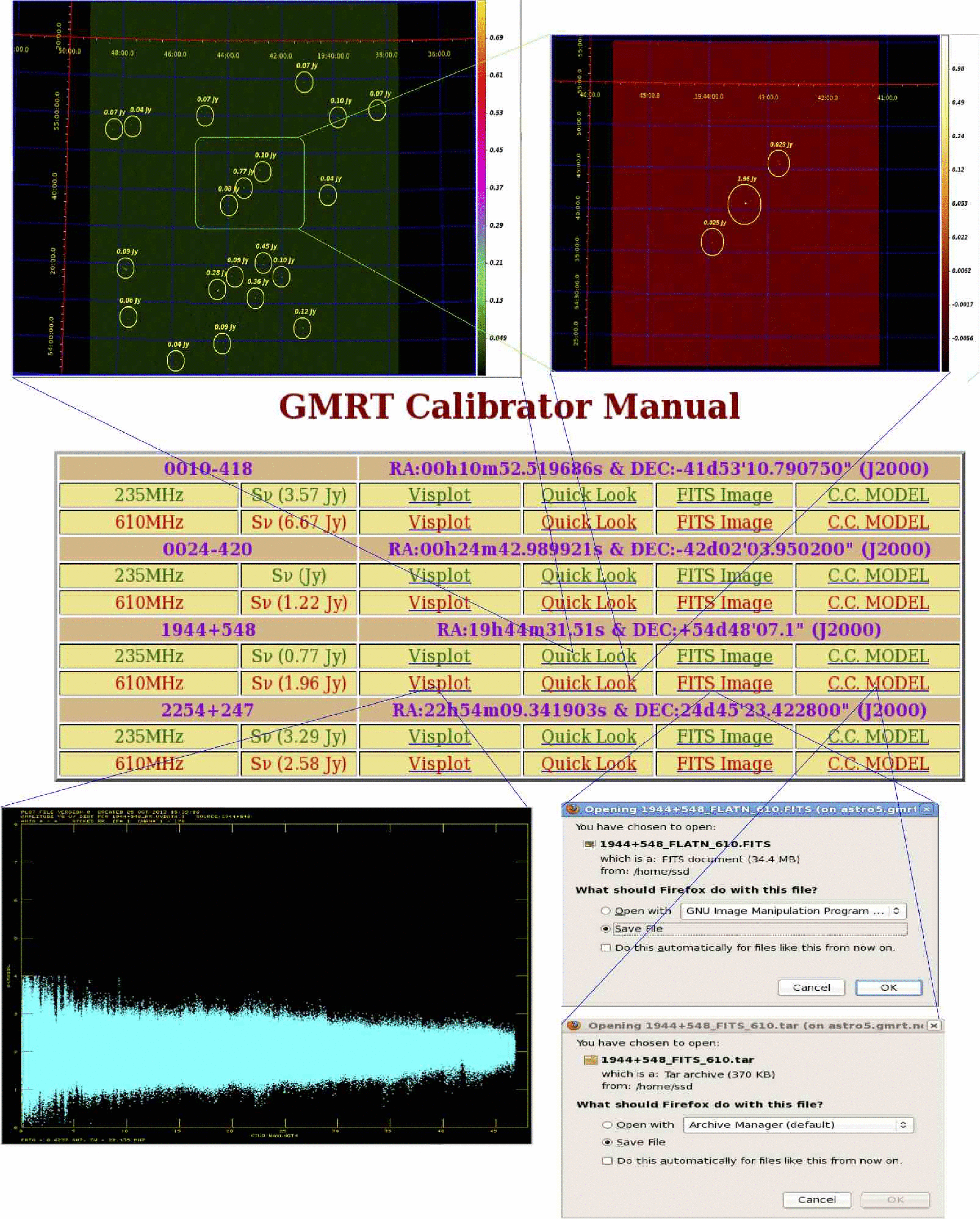}
\end{tabular}
\caption{A screen-shot of the GMRT calibrator manual from the online web-page at the
GMRT observatory.
Top-left panel and top-right are the images of a calibration source at 235 MHz and 610 MHz.
All the data products for a calibration source are also listed,
including 
coordinates ($\alpha$, $\delta$), observing frequency, flux density,
($u,v$) plot (or visplot), quick-look image of the field-of-view, \textsc{clean} restored model map
and \textsc{clean}-component model.}
\label{screenshot-file}
\end{center}
\end{figure*}

It is well known that
the technique of phase referencing permits the coherence time of the target source
data to be extended across the entire observation (many hours). This means that
the sensitivity of the observations continue to scale as
the square-root of the integration-time.
Here, we have introduced
and made available a database of phase calibration sources for the GMRT.
We provide their flux densities, models, ($u,v$) plots,
final deconvolved restored maps and \textsc{clean}-component lists/files
covering fields-of-view of $\sim$4 deg$^2$ and
$\sim$0.5 deg$^2$ at 235 MHz and 610 MHz, respectively,
for use in the \textsc{aips} and the \textsc{casa}
for all phase calibration sources in the database.
We also assign a quality factor for each of the calibration sources.
The data products are available through the GMRT observatory website.
A screen-shot of the GMRT calibrator manual from the online web-page
is shown in Fig.~\ref{screenshot-file}.
Additional findings from this study are:
\begin{itemize}
\item The distribution of these 45 phase calibration sources in the
sky is uniform with no visible large gaps or voids in the sky.
We used the gap statistics and the two-point correlation function
to demonstrate that the distribution of phase calibration sources are random
and the right ascension and sin($\pi/2 - {\rm declination}$) plane
is uniformly distributed.
Our ongoing efforts to increase the size of this database would
often provide a suitable choice of a phase calibration source that is within 20~deg to a
given target source.
\item Radio spectra have a variety of shapes, straight, curved and complex.
We find that the relative frequencies of different shapes are nearly equal
between straight, curved and complex shapes.
\item Quasars tend to exhibit flatter radio spectra as compared to the
radio galaxies or the unidentified sources.
\item Quasars are also known to be radio variable and hence possibly tend
to show complex spectra more frequently.
\item Radio galaxies tend to have
steeper spectra than either the quasars or the unidentified sources.
The steeper spectra are possibly either due to
the spectral ageing of synchrotron-emitting radio lobes
or due to the large redshift of distant galaxies
causing the shift of the spectrum to lower frequencies.
\end{itemize}

GMRT study of science target sources using calibration sources suitably chosen from the
database is an effective way to determine the phase calibration for the secondary
phase calibration sources and exploit fully the technique of phase-referencing.
We believe this is a valuable database 
(available at \url{http://gmrt.ncra.tifr.res.in})
in planning a GMRT observation, be it spectral-line or continuum or pulsar.
Our ongoing and future efforts will increase the size of this database,
thereby allowing the user to have a larger number of calibration sources.
The database at these two frequencies along with MSSS \citep{Healdetal}
and GLEAM-survey \citep{Waythetal} also help to interpolate
the information at 325 MHz.  325 MHz band is another GMRT frequency band with very
little information with regard to appropriate choices of suitable phase calibration sources.
This database would also be useful at the new low-frequency bands
of the upgraded GMRT\footnote{GMRT upgrade:
A major upgrade of several sub-systems of the GMRT has been initiated,
which will result in significant changes in almost all aspects of the GMRT
with the aim of significantly improving its capability and sensitivity.
Key features of the upgrade are near seamless frequency coverage from
125 MHz to 1450 MHz and instantaneous bandwidth of 400 MHz along with several matching
improvements in computing, receivers, servo and mechanical systems, electrical,
and civil structures.
The upgrade is nearing completion and the first phase of the upgraded system has already
been released to the astronomical community.} \citep{Gupta}.

\begin{acknowledgements}

We would like to thank the anonymous referees for their useful
suggestions and criticisms which helped improve the clarity of the paper.
We thank M.J. Hardcastle for a careful reading of this manuscript.
We thank the staff of the GMRT that made these observations possible.
GMRT is run by the National Centre for Radio
Astrophysics of the Tata Institute of Fundamental Research.
The VLA is operated by the US National Radio Astronomy Observatory which is
operated by Associated Universities, Inc., under cooperative agreement with
the National Science Foundation.  The National Radio Astronomy Observatory is
a facility of the National Science Foundation operated under cooperative
agreement by Associated Universities, Inc.
This research has made use of the NASA/IPAC Extragalactic Database (NED) which
is operated by the Jet Propulsion Laboratory, California Institute of
Technology, under contract with the National Aeronautics and Space
Administration.

\end{acknowledgements}

\end{document}